\documentclass[aps,prl,reprint]{revtex4-1}
\usepackage{epsf}
\usepackage{amsfonts}
\usepackage[fleqn]{amsmath}
\usepackage{amssymb,yfonts,mathrsfs,bbm}
\usepackage{graphicx}
\usepackage{multirow}
\usepackage{longtable}
\usepackage{amsmath}
\usepackage{color}
\usepackage{lineno}
\usepackage[colorlinks,
            linkcolor=red,
            anchorcolor=blue,
            citecolor=blue
            ]{hyperref}

\begin{document}

\title{Raman Forbidden Layer-Breathing Modes in Layered Semiconductor Materials Activated by Phonon and Optical Cavity Effects}
\author{Miao-Ling Lin$^{1,2}$, Jiang-Bin Wu$^{1,2}$, Xue-Lu Liu$^1$, Tao Liu$^{1,2}$, Rui Mei$^{1,2}$, Heng Wu$^{1,2}$, Shan Guan$^{1}$, Jia-Liang Xie$^{1,2}$, Jun-Wei Luo$^{1,2}$, Lin-Wang Wang$^{3}$, Andrea C. Ferrari$^{4}$, Ping-Heng Tan$^{1,2}$}
\email{phtan@semi.ac.cn}
\affiliation{$^1$ State Key Laboratory of Semiconductor Physics and Chip Technologies, Institute of Semiconductors, Chinese Academy of Sciences, Beijing 100083, China}
\affiliation{$^2$ Center of Materials Science and Optoelectronics Engineering, University of Chinese Academy of Sciences, Beijing, 100049, China}
\affiliation{$^3$ State Key Laboratory of Optoelectronic Materials and Devices, Institute of Semiconductors, Chinese Academy of Sciences, Beijing 100083, China}
\affiliation{$^4$ Cambridge Graphene Centre, University of Cambridge, 9 JJ Thomson Avenue, Cambridge CB3 0FA, UK}
%\linenumbers

\begin{abstract}
We report Raman forbidden layer-breathing modes (LBMs) in layered semiconductor materials (LSMs). The intensity distribution of all observed LBMs depends on layer number, incident light wavelength and refractive index mismatch between LSM and underlying substrate. These results are understood by a Raman scattering theory via the proposed spatial interference model, where the naturally occurring optical and phonon cavities in LSMs enable spatially coherent photon-phonon coupling mediated by the corresponding one-dimensional periodic electronic states. Our work reveals the spatial coherence of photon and phonon fields on the phonon excitation via photon/phonon cavity engineering.
%\textbf{Keywords:} Phonon cavity effects, Optical cavity effects, Layered semiconductor materials, Layer-breathing mode, Tunable phonon excitation
\end{abstract}
\maketitle

%\section{INTRODUCTION}
Raman scattering has been a ubiquitous tool for probing elementary excitations\cite{Cardona-1983,Gadelha-nat-2021,li-prl-2020} (e.g., phonons) and electron (exciton)-photon/phonon (e-pht/e-phn) interactions\cite{Eliel-nc-2018,Zhang-nc-2022,QHTan-2023-NC} in both bulk and nanoscale materials. In the quantum picture of Raman scattering, incident photons first excite a set of intermediate electronic states, which then generate or absorb phonons and give rise to energy-shifted scattered photons.\cite{Cardona-1983,chen-nat-2011} The intermediate electronic excitations are pivotal as quantum pathways in Raman scattering\cite{lin-npho-2010,chen-nat-2011,miranda-nl-2017} and determine the e-pht and e-phn interaction matrix elements\cite{Cardona-1983,Carvalho-prl-2015,Eliel-nc-2018,QHTan-2023-NC}. Traditionally, these can be evaluated by making a multipole expansion\cite{Cardona-1983}, based on the premise that the wavelength of light is large compared with atomic dimensions. Usually, only the first term of the multipole expansion, i.e., electric dipole, is retained\cite{Cardona-1983}, defining the Raman tensor based on group symmetry analysis to determine the polarization selection rules.\cite{Cardona-1983,Loudon-ap-1964} An interlayer bond polarizability model (IBPM) within this approximation was also developed to understand the Raman intensity of interlayer phonons in ultrathin layered materials (LMs)\cite{luo2015stacking,liang2017interlayer}, and in LM heterostructures showing cross-dimensional e-phn coupling\cite{Lin-2019-NC}. In this case, the e-pht matrix element is considered independent of the photon wavevector.

\begin{figure*}[]
\centerline{\includegraphics[width=180mm,clip]{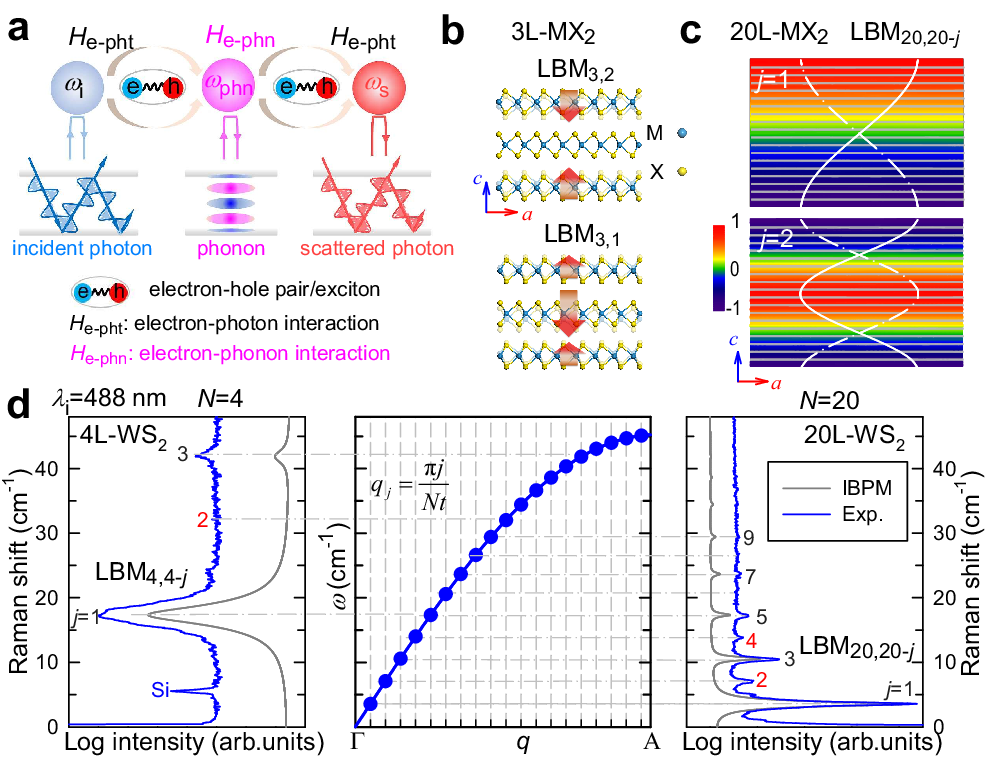}}
\caption{\textbf{a}. Raman scattering modulated by photon/phonon fields overlapping in a cavity. \textbf{b}. Displacements of LBMs in 3L-MX$_2$, where the arrow length represents the vibration amplitude. \textbf{c}. Side view of displacement field profiles of LBM$_{20,20-j}$ ($j$=1,2) in 20L-$MX_2$, where the color code represents the amplitude. \textbf{d}. $\omega-q$ along $\Gamma-A$ direction with $q$-dependent Pos(LBM$_{N,N-j}$) in 4L- and 20L-WS$_2$, and experimental (blue) and IBPM predicted (gray) Raman spectra of the corresponding LBMs. }\label{Fig1}
\end{figure*}

If light wavelength and dimension of phonon displacement field are comparable, the premise for multipole expansion fails, leading to the breakdown of the Raman selection rules based on symmetry analysis. It is a challenge to achieve matching between light wavelength and dimension of phonon displacement fields for Raman scattering in bulk solids. However, LMs can act as phonon cavity \cite{Zalalutdinov-nc-2021,ge-srep-2014,Soubelet-nanoscale-2017} by generating standing waves, enabling wavelength (and wavevector) matching between photon field and the phonon field standing waves by adjusting LM thickness and dielectric environment. LMs can also generate Fabry-P\'{e}rot optical cavity with photon field redistribution for enhanced light-matter interaction\cite{xu-npho-2022,Zhang-np-2022}, which could induce a change in photon-phonon (pht-phn) coupling. These cavity effects could result in new Raman selection rules.

Here, we show that layered semiconductor materials (LSMs) with specific thicknesses can act as naturally occurring optical cavities and induce spatial variations in the photon field inside LSMs, and phonon cavity to match the photon wavevector with quantized standing-wave vectors of layer-breathing phonons along the out-of-plane axis. The two effects result in spatially modulated e-pht and e-phn interactions and the observation of Raman forbidden layer-breathing modes (LBMs). A spatial interference model (SIM) of pht-phn coupling that integrates e-pht and e-phn interactions is presented to quantitatively describe the intensity of Raman-inactive LBMs dependent on LBM standing-wave vectors, LSM layer number, excitation wavelength and the underlying substrate. This novel Raman scattering theory of pht-phn coupling goes beyond the traditional electric dipole approximation in Raman tensor theory.

We consider the interaction diagram in a Raman scattering event involving one phonon excitation (Fig.~\ref{Fig1}a), where $\omega_{\rm i}$, $\omega_{\rm s}$, and $\omega_{\rm phn}$ are the frequencies of incident (i), scattered (s) photons, and phonon, respectively. The interactions between incident/scattered photons and phonons are mediated by electron-hole (e-h) pairs\cite{Cardona-1983} or excitons\cite{Sven-sciadv-2020}, where e-pht and e-phn interactions are characterized by the corresponding matrix elements $M_{\rm e-pht(i/s)}$ and $M_{\rm e-phn}$ \cite{Cardona-1983}. The output in this scattering event is determined by a third-order perturbation process\cite{Cardona-1983,Mlayah-prb-2007}:

\begin{equation}
\begin{aligned}
I \propto \left | \sum_{e,e'}\frac{ M_{\rm e-pht(i)}  M_{\rm e-phn} M_{\rm e-pht(s)} }{(\hbar\omega_{\rm i}-\epsilon_{\rm e-h}+i\gamma)(\hbar\omega_{\rm s}-\epsilon_{\rm e'-h}+i\gamma )} \right |^{2}
\end{aligned}
\label{eq:inni}
\end{equation}

\noindent where $h$, $e$, and $e'$ label the states of photoexcited hole, photoexcited electron, and scattered electron, respectively, $\omega_{\rm s}=\omega_{\rm i}\pm\omega_{\rm phn}$, $\epsilon_{\rm e-h}$ and $\epsilon_{\rm e'-h}$ are the energies of electronic transitions, and $\gamma$ is the homogeneous linewidth of the electronic transition\cite{Cardona-1983,Mlayah-prb-2007}. The Raman intensity is sensitive to the pht-phn coupling, $i.e.$, $ M_{\rm e-pht(i)}  M_{\rm e-phn} M_{\rm e-pht(s)} $, and the detuning of the involved photon from inherent optical resonances\cite{Cardona-1983,chen-nat-2011,Ferrari-nn-2013,corro-acsnano-2014,Carvalho-prl-2015,kim-acsnano-2016}. Once incident/scattered photon and phonon are confined and overlap in a cavity, both $M_{\rm e-pht}$ and $M_{\rm e-phn}$ can exhibit spatial variations within the cavity, resulting in constructive or destructive interference of pht-phn coupling in real space.

Here we use LSM flakes with layer number $N>1$, such as $MX_2$ (M=Mo, W and X=S, Se) ($N$L-$MX_2$). These have $N-1$ phonon modes involving the relative motions of atomic planes along the out-of-plane ($c$) axis, i.e., LBM$_{N,N-j}$ ($j$=1,2...$N$-1)\cite{Tan-2012-NM,zhang2013raman,zhao-nl-2013,Ferrari-nn-2013,pizzi-acsnano-2021}, as depicted by LBM$_{3,3-j}$ ($j$=1,2) of 3L-$MX_2$ in Fig.~\ref{Fig1}b.  According to the linear chain model\cite{Tan-2012-NM,zhang2013raman,zhao-nl-2013,pizzi-acsnano-2021}, the atomic plane displacements are given by $u_{\rm LBM}(q_j,z)=(e^{{\rm i}q_j z}+e^{-{\rm i}q_j z})/2$, where $q_j={\pi j}/{(Nt)}=j\delta q$ ($j$=1,2...$N$-1) with $\delta q$ being the $q_j$ difference between two adjacent LBMs, $z=(2n-1)t/2$ with $n$ the layer index and $t$ the thickness of monolayer. Thus, LBMs can be considered as phonon cavity modes showing standing-wave feature with wavelength $\lambda_{{\rm LBM}_{N,N-j}}=2Nt/j$, i.e, the superposition of two one-dimensional (1d) counterpropagating plane-wave components (out-of-plane), labelled $+q_j(-q_j)$ for the components propagating away (towards) the sample surface, as exemplified by the displacement field profiles of LBM$_{20,20-j}$ ($j=1,2$) for 20L-$MX_2$ (color contour and white curves in Fig.~\ref{Fig1}c). The quantized $q_j$ along the $c$ axis corresponds to the standing-wave vector of LBM displacement fields in LSMs. This phonon cavity characteristic for LBMs in LSMs provide a platform to enable wavevector matching between photon and the standing wave of phonons by tuning $N$. The frequencies of the $N-1$ LBMs are also dependent on $q_j$, and can be derived from the linear chain model\cite{Tan-2012-NM,zhang2013raman,zhao-nl-2013,pizzi-acsnano-2021} as $\text{Pos}(\text{LBM}_{N,N-j})=\text{Pos}(\text{LBM}_\infty)\sin(q_{j}t/2)$, where $\text{Pos}(\text{LBM}_{\infty})$ represents the peak position of the LBM in a bulk LSM. $\text{Pos}(\text{LBM}_{N,N-j})$ can be also expressed by the $\omega-q$ of the longitudinal acoustic phonons in bulk $MX_2$ along $\Gamma-A$ direction ($q_{\rm A}$=$(0,0,\pi/t)$)\cite{karssemeijer2011phonons}, in which the confinement of $N$L-$MX_2$ in the $c$ axis limits $q_j$ of LBM$_{N,N-j}$ to integral multiples of $\pi/Nt$, as demonstrated for 4L- and 20L-WS$_2$ in Fig.~\ref{Fig1}d.

To probe the LBMs in WS$_2$ flakes, we use an incident photon energy of 2.54 eV (Section I and Fig. S1 of the Supplementary Materials (SM)\cite{SM-2024}), resonant with the C exciton energy $\sim$2.6 eV\cite{zhao2012evolution,tan2017observation}. The experimental LBM$_{N,N-j}$ intensity distribution of 4L-WS$_2$ (Fig.~\ref{Fig1}d) can be understood by the IBPM (Section II of the SM\cite{SM-2024}), with odd LBM$_{N,N-j}$ ($j$ is odd) peaks and absence of even LBM$_{N,N-j}$ ($j$ is even, Fig. S2 of the SM\cite{SM-2024}). However, for 20L-WS$_2$, significant differences emerge, particularly with the rising intensity of even LBM$_{N,N-j}$ peaks, predicted to be Raman-inactive by the IBPM  (Fig.~\ref{Fig1}d) and symmetry analysis. 20L-WS$_2$ exhibits much smaller $q_j$ (0.25 nm$^{-1}$ when $j=1$) than that (1.27 nm$^{-1}$ when $j=1$) in 4L-WS$_2$. The former is comparable with the change of incident light wavevector ($\Delta k$=$2k_{\rm i}$=0.13 nm$^{-1}$, $k_{\rm i}={2\pi \tilde{n}_1}/{\lambda_{\rm i}}$ is the wavevector at incident laser wavelength $\lambda_{\rm i}$ within a LSM with refractive index $\tilde{n}_1$) in back scattering configuration. This suggests a novel mechanism ruling the observation of forbidden even LBM$_{N,N-j}$, due to the wavevector matching between incident/scattered photon and a standing wave of LBM phonon within 20L-WS$_2$ cavity.

For specific $\hbar\omega_{\rm i}$ (or $\lambda_{\rm i}$), the Raman intensity of LBMs in LSMs is usually determined by $M_{\rm e-pht(i)} M_{\rm e-phn} M_{\rm e-pht(s)}$ in Eq.(\ref{eq:inni}). We first consider the e$-$phn interaction term $M_{\rm e-phn}$. $\hbar\omega_{\rm i}$=2.54 eV can resonantly excite e-h pairs close to $\Gamma$ point of Brillouin zone related to the C exciton of WS$_2$ flakes, confined within each layer\cite{qiu-prl-2013} (Section III and Fig. S3 of SM\cite{SM-2024}) and form an ensemble of confined electronic states ($i.e.$, 1d periodic electronic states\cite{qiu-prl-2013,Takeyama-nl-2021}). The standing-wave nature of LBMs generates a lattice dilation along the $c$ axis. Thus, $M_{\rm e-phn}$ between delocalized LBMs and the 1d periodic electronic states can be expressed by the deformation-potential interaction\cite{kittel1976introduction,Mlayah-prb-2007} as (Section III of SM\cite{SM-2024}):
\begin{equation}
\begin{aligned}
&M_{\rm e-phn} \propto \frac{\frac{1}{2} \pm \frac{1}{2} + n(q_j)}{\omega_{\rm phn}} \int |\Psi(z)|^2 m_{\rm ph}(q_jz) dz
\end{aligned}
\end{equation}
\noindent where $\Psi(z)$ is the 1d periodic electronic wave function, $m_{\rm ph}(q_j,z)$ = $e^{{\rm i} q_j z} - e^{-{\rm i} q_j z}$ gives the lattice dilation along the $c$ axis, $n(q_j)$ is the Bose-Einstein population factor, and the $+$ ($-$) sign stands for Stokes (anti-Stokes) Raman scattering. The layer displacements of LBMs lead to significant spatial modulation of $M_{\rm e-phn}$, which depends on $q_j$. Within the electric dipole approximation, $M_{\rm e-pht(i/s)}$ is considered as constant with polarization direction of the incident (scattered) light. In this case, the LBM intensity is proportional to $|M_{\rm e-phn}|^2$. The corresponding calculated LBM spectrum (Fig. S4 of SM\cite{SM-2024}) does not have even LBMs for 20L-WS$_2$, in agreement with the IBPM, but in contrast with experiments. This suggests we need to go beyond the electric dipole approximation for the e$-$pht coupling term $M_{\rm e-pht(i/s)}$.

\begin{figure*}[]
\centerline{\includegraphics[width=180mm,clip]{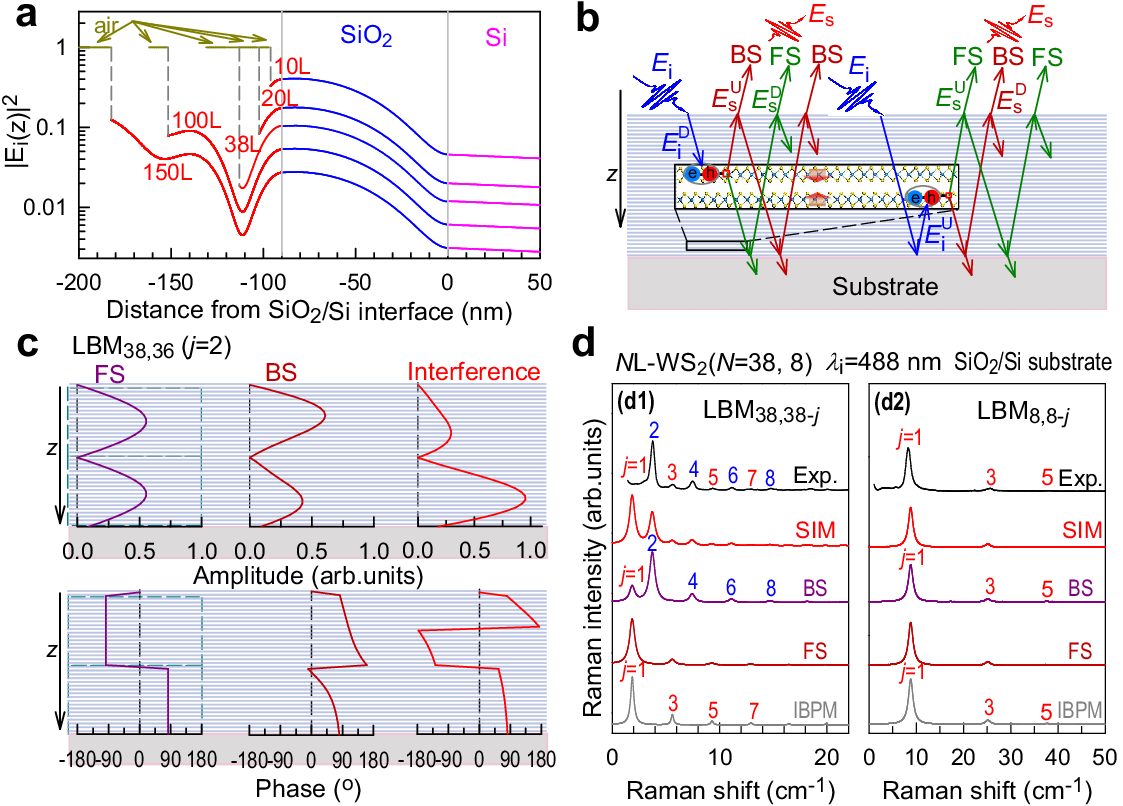}}
\caption{\textbf{a}. $|E_{i}(z)|^2$ in the $N$L-WS$_2$/SiO$_2$/Si structures.  \textbf{b}. Propagation path of incident (blue) and scattered (red) photons in $MX_2$ flakes. Two pathways of incident and scattered photons, toward (up,U) and away from (down,D) the top surface of $MX_2$, result in internal FS and BS geometries. \textbf{c}. Spatial variation of amplitude and phase of FS, BS and FS-BS interfering pht-phn coupling matrixes for LBM$_{38,36}$ of 38L-WS$_2$. \textbf{d}. LBM spectra of (d1) 38L-WS$_2$ and (d2) 8L-WS$_2$ on SiO$_2$/Si for FS and BS components and FS-BS interference calculated by SIM, along with IBPM-simulated and experimental (Exp.) spectra.}\label{Fig2}
\end{figure*}

We now consider the detailed incident/scattered photon propagation in the air/LSM/substrate dielectric multilayers for Raman scattering of LBMs. The refractive index mismatch($\Delta_{\tilde{n}} = (\tilde{n}_{\mu} - \tilde{n}_{\nu}) / (\tilde{n}_{\mu} + \tilde{n}_{\nu})$, with $\tilde{n}_{\mu}$ and $\tilde{n}_{\nu}$ the complex refractive indexes of the adjacent media) between LSM, air, and substrate can lead to partial reflection of light at the air/LSM and LSM/substrate interfaces, resulting in an optical cavity effect. This leads to a considerable spatial modulation in the modulus square of the electric field for the incident laser inside LSMs with thickness up to 150L, as shown in Fig.~\ref{Fig2}a, due to the superposition of forward (down) and backward (up) propagating optical waves (whose electric field components are denoted as $E^{\rm D}_{\rm i}$ and $E^{\rm U}_{\rm i}$, respectively), as shown in Fig.~\ref{Fig2}b. The scattered optical wave generated by the recombination of scattered e-h pairs by LBMs also has two components propagating toward to (up) and away from (down) the air/LSM interface, $i.e.$, $E^{\rm U}_{\rm s}$ and $E^{\rm D}_{\rm s}$. The combination of incident and scattered electric field components can lead to two internal scattering geometries within LSM flakes, i.e., forward (FS) and backward scattering (BS). Accordingly, the e-pht coupling ($m_{\rm e-pht(i)}(z)m_{\rm e-pht(s)}(z)$) at position $z$ is derived by the product of incident and scattered electric field components (Section IV of SM\cite{SM-2024}):
\begin{equation}
\begin{aligned}
&m_{\rm e-pht(i)}(z)m_{\rm e-pht(s)}(z)\propto(E^{\rm U}_{\rm i}+E^{\rm D}_{\rm i})(E^{\rm U}_{\rm s}+E^{\rm D}_{\rm s})\\
&=t_{01}t_{10}\frac{2r_{12}e^{2{\rm i}k_{\rm i}Nt}+[e^{2{\rm i}k_{\rm i}z}+r^2_{12}e^{2{\rm i}k_{\rm i}(2Nt-z)}]}{(1-r_{12}r_{10}e^{2{\rm i}k_{\rm i}Nt})^2}\\
&=m_{\rm e-pht}^{\rm FS}(2k_{\rm i})+m_{\rm e-pht}^{\rm BS}(2k_{\rm i}z)
\end{aligned}
\label{eq:fsbs}
\end{equation}
\noindent with $t_{\mu\nu}$ and $r_{\mu\nu}$ the amplitude transmission and reflection coefficients from medium $\mu$ to $\nu$, respectively, and $m_{\rm e-pht}^{\rm FS}(2k_{\rm i})$ and $m_{\rm e-pht}^{\rm BS}(2k_{\rm i}z)$ representing the e-pht coupling of the FS and BS components.

The spatial distribution of pht-phn coupling along $c$ axis, including e-pht and e-phn interactions mediated by the electronic states, can be written as $[m_{\rm e-pht}^{\rm FS}(2k_{\rm i})+ m_{\rm e-pht}^{\rm BS}(2k_{\rm i}z)]m_{\rm e-phn}(q_jz)$ (Section V of SM\cite{SM-2024}). Because of the localization of electronic states within the layer, a coherent summation of spatially modulated pht-phn coupling for the whole LSM leads to an variation of Raman intensity ($I$) as a function of $q_j$ (or Pos(${\rm LBM}_{N,N-j}$)):
\begin{equation}
\begin{aligned}
I(q_j)&\sim\bigg|\sum_{n=0}^{N-1}[m_{\rm e-pht}^{\rm FS}(2k_{\rm i})+m_{\rm e-pht}^{\rm BS}(2k_{\rm i}n)]m_{\rm e-phn}(q_jn)\bigg|^2\\
&=\bigg|\sum_{n=0}^{N-1}\{S_{\rm FS}(2k_{\rm i},q_jn)+S_{\rm BS}[(2k_{\rm i} \pm q_j)n]\}\bigg|^2,
\end{aligned}
\label{eq:sinc3}
\end{equation}
\noindent where the $S_{\rm FS}(2k_{\rm i}, q_jn)$ ($S_{\rm BS}[(2k_{\rm i} \pm q_j)n]$) is the pht-phn coupling of the FS (BS) component in the $n^{th}$ LSM layer. We denote the above model described by Eq. (\ref{eq:sinc3}) as the SIM. This shows how the spatially coherent pht-phn coupling in the phonon and optical cavities of LSMs determines the Raman intensity of the corresponding LBMs.

\begin{figure*}[]
\centerline{\includegraphics[width=180mm,clip]{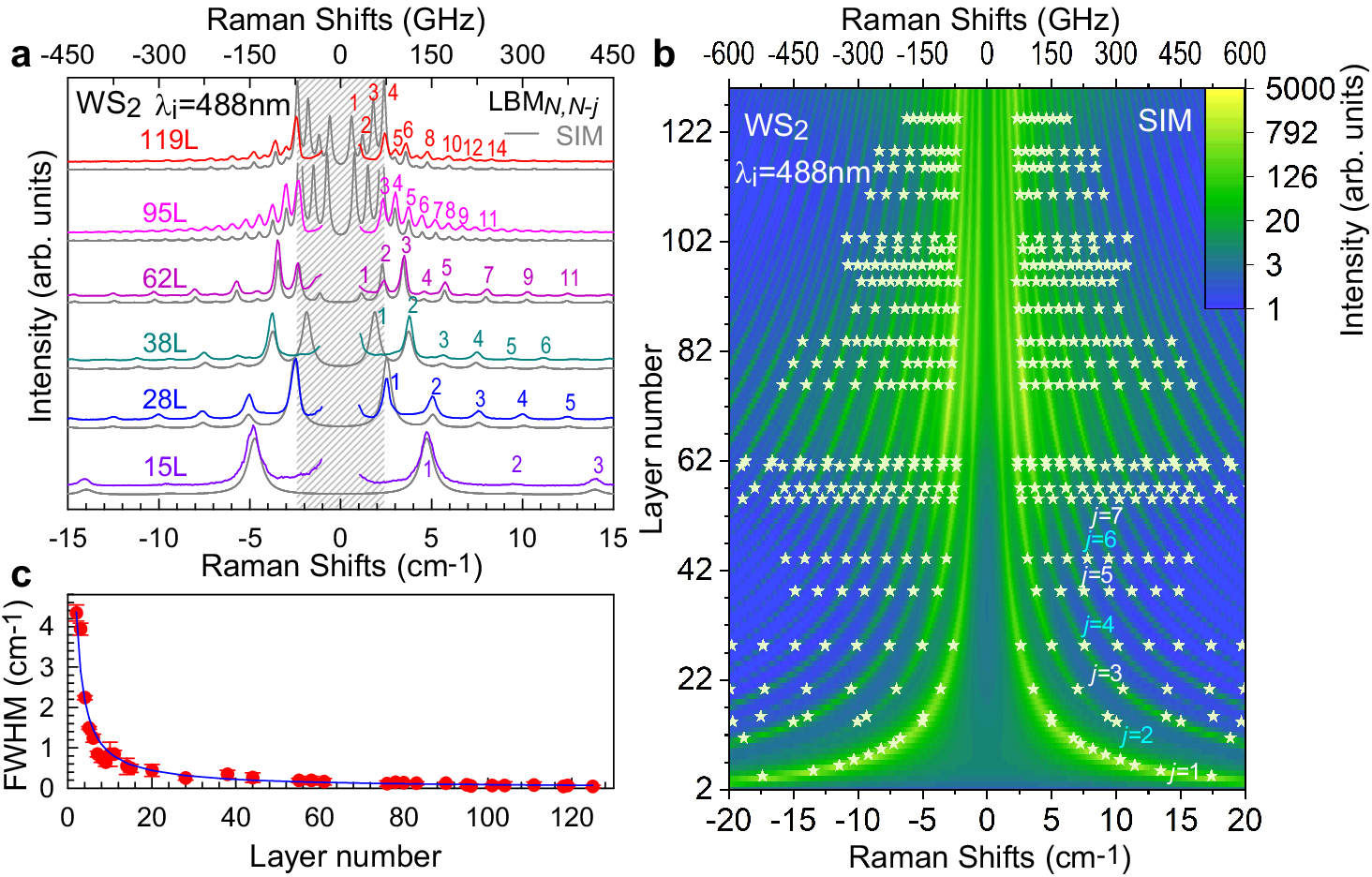}}
\caption{\textbf{a}. SIM-calculated (gray) and experimental (colored) LBM spectra. The stripe-pattered shade is the blind spectral range of our Raman setup. \textbf{b}. SIM-calculated phonon spectra in $N$L-WS$_2$ ($N$=2-130). Stars are experimental data. \textbf{c}. $1/N$-dependent average FWHM, where 0.16 cm$^{-1}$ is subtracted from FWHM to account for the system broadening.} \label{Fig3}
\end{figure*}

We first analyze the LBM intensity distribution of thick LSMs (e.g., 38L-WS$_2$) based on SIM using the corresponding complex refractive indices: $5.2+1.1$i\cite{li2014measurement} for WS$_2$, $1.4629$\cite{Palik-2002} for SiO$_2$, and $4.3606+0.0868$i\cite{Palik-2002} for Si. The calculated spatial variations in amplitude and phase of $S_{\rm FS/BS}$ along the $c$ axis for LBM$_{38,36}$ ($j$=2) of 38L-WS$_2$ on SiO$_2$/Si are depicted in Fig.~\ref{Fig2}c. The amplitude and phase of $S_{\rm FS}$ synchronously oscillate in real space related to $q_j$ (Section V of SM\cite{SM-2024}). For even $j$ (e.g., LBM$_{38,36}$, $j$=2), the phase varies alternately between positive and negative values, with a shift of $\pi$ (Fig.~\ref{Fig2}c and Fig. S5 of SM\cite{SM-2024}). A coherent summation of $S_{\rm FS}(2k_{\rm i}, q_jn)$ for the whole LSM could result in zero intensity for even LBMs, due to spatially destructive interference (Fig. S5 of SM\cite{SM-2024}), and finite Raman intensity for odd LBMs because of incompletely destructive interference (Fig. S6 of SM\cite{SM-2024}). Conversely, the amplitude and phase of $S_{\rm BS}$ is a function of $2k_{\rm i} \pm q_j$, and the phase does not always vary alternately between positive and negative values (Fig.~\ref{Fig2}c, Fig. S5 and Fig. S6 of SM\cite{SM-2024}). Thus, a coherent summation of $S_{\rm BS}[(2k_{\rm i} \pm q_j)n]$ for the whole LSM leads to the LBM intensity varying with the difference between $2k_{\rm i}$ and $q_j$, where the even LBM intensity is not always zero. By considering the general Lorentzian lineshape of LBMs and the weighted intensity, we calculate the LBM Raman spectra from FS/BS components and FS-BS interference (Fig.~\ref{Fig2}d1). When just considering the FS component, $I(q_j)$ exhibits a series of zeroes at even $j$, leading to the observation of only odd LBMs, aligning with the IBPM predictions. However, if one just considers the BS component, even LBMs can display significantly higher intensities than the adjacent odd ones. The interference of FS and BS components leads to the excitation of both odd and even LBMs in the Raman spectrum, and the corresponding calculated LBM spectrum agrees with the experimental one (Fig.~\ref{Fig2}d1).

For thin LSMs, e.g., WS$_2$ flakes with small $N$ (e.g., $N<10$), the condition $2k_{\rm i} \ll \Delta q$, i.e., $|q_j| \sim |2k_{\rm i} \pm q_j|$, leads to $S_{\rm BS}$ being close to $S_{\rm FS}$. This similarity results in that the spatial amplitude and phase of the FS-BS interference closely approximate those of $S_{\rm FS}$. Therefore, Raman spectra from FS-BS interference and BS component are similar to that of the FS component, such as the case of 8L-WS$_2$ in Fig.~\ref{Fig2}d2. As $N$ increases, the mismatch between $q_j$ and $2k_{\rm i} \pm q_j$ enlarges, and the conditions for destructive interference cease to hold, allowing the even modes to emerge, as discussed above for 38L-WS$_2$.

\begin{figure*}[]
\centerline{\includegraphics[width=180mm,clip]{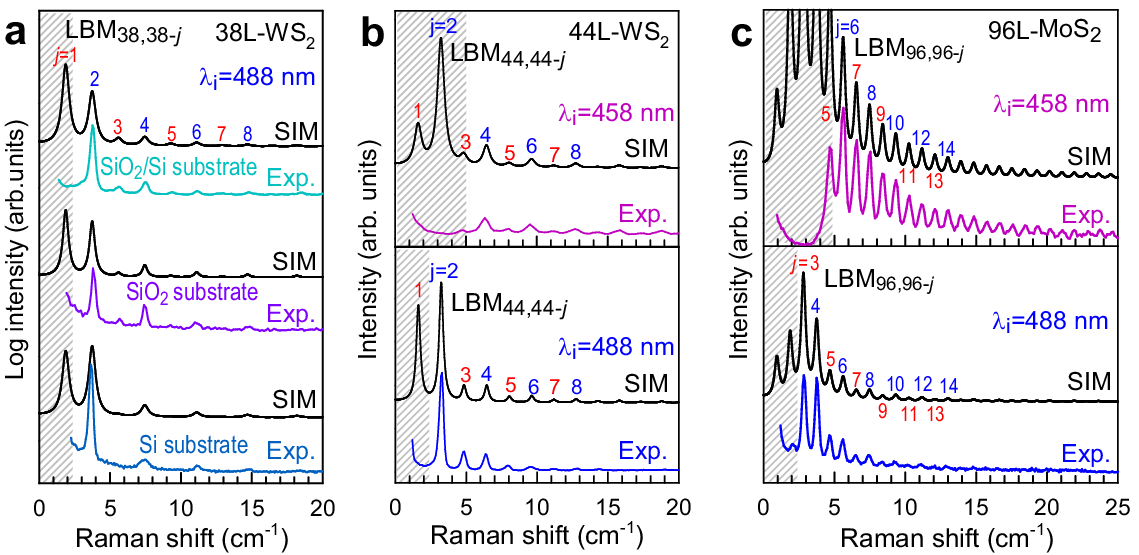}}
\caption{\textbf{a}. Experimental (Exp., colored) and SIM-calculated (black) LBM spectra of 38L-WS$_2$ on different substrates, $\lambda_{\rm i}$=488 nm, and those for (\textbf{b}) 44L-WS$_2$ and (\textbf{c}) 96L-MoS$_2$ with $\lambda_{\rm i}$=458 nm and 488 nm.} \label{Fig4}
\end{figure*}

To clearly reveal the evolution of the LBM emission with varying LSM thickness, we measured LBM spectra (Fig.\ref{Fig3}a,b) of $N$L-WS$_2$ flakes on SiO$_2$/Si with $N$ varying from 2 to 130 under resonant excitation of the C exciton. The calculated LBM spectra concur with the experimental results. The intensities of even LBMs relative to that of adjacent odd LBMs are significantly enhanced in the specific $N$ ranges, as demonstrated in Fig.\ref{Fig3}b. E.g., with $N = 10 \sim 30$, the odd LBM$_{N,N-j}$ branches exhibit stronger intensities than the adjacent even LBM$_{N,N-j-1}$ branches, whereas for $N = 30 \sim 60$, the intensities of the observed even LBM$_{N,N-j}$ branches surpasses those of adjacent odd LBM$_{N,N-j-1}$ ones.
%\textcolor{red}{, where the Raman intensity of the forbidden even LBMs gradually change as a function of $N$}

Figures \ref{Fig3}a,b demonstrate that the frequency difference between adjacent LBMs of $N$L-WS$_2$ decreases as $N$ increases. The average full width at half maximum (FWHM) of the LBMs also decreases with increasing $N$, showing a $1/N$ dependence (Fig.\ref{Fig3}c, blue line). This can be ascribed to the $q_j$ confinement of standing-wave LBMs along the $c$ axis. The uncertainty of $q$ [$\Delta q$] can be estimated by the Heisenberg uncertainty principle, i.e., $\Delta x \Delta p \sim \hbar/2$, where $\Delta x=Nt$ and $\Delta p=\hbar\Delta q$. According to the Pos(${\rm LBM}$)$-q_j$ relation, FWHM(${\rm LBM})\propto (t/2){\rm cos}(q_j t/2)  \times \Delta q \propto 1/N$.
% in the large $N$ ($N>10$, ${\rm cos}(q_1 t/2)\sim 1$) limit
%%\textcolor{red}{FWHM of $N$L-WS$_2$ is about 1$\sim$4 cm$^{–1}$ for $N<$10, but can be as small as 0.1 cm$^{–1}$ when $N$ approaches 100 after excluding the system broadening.}

SIM also implies that the intensity distribution of all the observed LBMs is sensitive to the optical cavity effect, which should be significantly dependent on the refractive index mismatch between LSM and its underlying substrates. As a check, we measure the Raman spectra (Fig.~\ref{Fig4}a) of 38L-WS$_2$ flakes on 90 nm-SiO$_2$/Si, bare SiO$_2$, and Si. A significant refractive index mismatch ($\Delta{\tilde{n}} \approx 0.57 + 0.07$i) between WS$_2$ and SiO$_2$ enhances the optical cavity effect, boosting the FS component (odd LBMs, Figs.\ref{Fig2}b,d1) contributions to LBM spectra on 90 nm-SiO$_2$/Si and bare SiO$_2$. However, when 38L-WS$_2$ flakes is deposited on Si, due to the close refractive indexes of WS$_2$ and Si, the optical cavity effect at the 38L-WS$_2$/Si interface is weak, and the FS component contribution is small. Thus, odd LBMs are absent in the corresponding Raman spectrum with dominant even LBMs. The measured LBM spectra for 38L-WS$_2$ on different substrates agree with the SIM predictions. These results demonstrate the possibility to tune LBM emissions by varying the refractive index mismatch between LSM and its substrate.

Due to the importance of wavevector matching between the photon and standing wave of phonons in ruling the LBM Raman intensity, it is expected that the LBM emission will be sensitive to $\lambda_{\rm i}$. Figure \ref{Fig4}b presents the experimental and calculated Raman spectra of LBMs in 44L-WS$_2$ excited at 488 nm and 458 nm, under resonant excitation of the C exciton\cite{tan2017observation}. The different $k_{\rm i}$ lead to varying mismatches between $q_j$ and $2k_{\rm i} \pm q_j$, resulting in distinctly different interference effects at these $\lambda_{\rm i}$. Compared to $\lambda_{\rm i}$=488 nm, the intensities of odd LBMs are weaker than those of adjacent even LBMs under $\lambda_{\rm i}$=458 nm in 44L-WS$_2$. Similar $\lambda_{\rm i}-$dependent LBM spectra can be observed in other LSMs, such as MoS$_2$, MoSe$_2$, and MoTe$_2$, as illustrated in Fig.~\ref{Fig4}c and Fig. S7 (Section VI of SM\cite{SM-2024}). This supports the general applicability of the SIM, which accounts for the spatially coherent pht-phn coupling for LBMs in LSM cavities.

In conclusion, we reported the observation of forbidden LBM emission in LSMs, driven by spatially coherent coupling of the photon field propagating along the $c$ axis and the phonon field to the 1d periodic electronic states, which can be well understood by a Raman scattering theory via the proposed SIM. The Raman intensity distribution of all the observed LBMs can be tuned by varying the wavevector matching between photons and standing wave of phonons with LSM thickness, excitation wavelength, and refractive index mismatch between LSM and underlying substrate.

\vspace*{5mm}
%\noindent {\bf \large References}

%\vspace*{5mm}
%\noindent {\bf \large Acknowledgements}

\noindent We acknowledge the support from the National Key Research and Development Program of China (Grant No. 2023YFA1407000), the Strategic Priority Research Program of CAS (Grant No. XDB0460000), National Natural Science Foundation of China (Grant Nos. 12322401, 12127807 and 12393832), CAS Key Research Program of Frontier Sciences (Grant No. ZDBS-LY-SLH004), Beijing Nova Program (Grant No. 20230484301), Youth Innovation Promotion Association, Chinese Academy of Sciences (No. 2023125), Science Foundation of the Chinese Academy of Sciences (Grant No. JCPYJJ-22), CAS Project for Young Scientists in Basic Research (YSBR-026), EU Graphene and Quantum Flagships, ERC Grants Hetero2D, GIPT, EU Grants GRAP-X, CHARM, EPSRC Grants EP/K01711X/1, EP/K017144/1, EP/N010345/1, EP/L016087/1, EP/V000055/1, EP/X015742/1.

\bibliographystyle{naturemag}

\end{document}

%\title{Tunable interlayer phonon emission in two-dimensional semicondutors}
\title{Supplementary Materials for \\
Raman Forbidden Layer-breathing Modes in Layered Semiconductor Materials Activated by Phonon and Optical Cavity Effects}

\author{Miao-Ling Lin$^{1,2}$, Jiang-Bin Wu$^{1,2}$, Xue-Lu Liu$^1$, Tao Liu$^{1,2}$, Rui Mei$^{1,2}$, Heng Wu$^{1,2}$, Shan Guan$^{1}$, Jia-Liang Xie$^{1,2}$, Jun-Wei Luo$^{1,2}$, Lin-Wang Wang$^{3}$, Andrea C. Ferrari$^{4}$, Ping-Heng Tan$^{1,2}$}
\email{phtan@semi.ac.cn}
\affiliation{$^1$ State Key Laboratory of Semiconductor Physics and Chip Technologies, Institute of Semiconductors, Chinese Academy of Sciences, Beijing 100083, China}
\affiliation{$^2$ Center of Materials Science and Optoelectronics Engineering \& CAS Center of Excellence in Topological Quantum Computation, University of Chinese Academy of Sciences, Beijing, 100049, China}
\affiliation{$^3$ State Key Laboratory of Optoelectronic Materials and Devices, Institute of Semiconductors, Chinese Academy of Sciences, Beijing 100083, China}
\affiliation{$^4$ Cambridge Graphene Centre, University of Cambridge, 9 JJ Thomson Avenue, Cambridge CB3 0FA, United Kingdom}
%\linenumbers
\begin{abstract}

\end{abstract}
\maketitle

\section{Methods}
\textbf{Sample preparation:} Multilayer transitional metal dichalcogenide flakes, denoted as $MX_2$ ($M$=Mo,W and $X$=S, Se, Te) are mechanically exfoliated from the bulk counterparts (2D semiconductors USA) onto 90 nm-SiO$_2$/Si, SiO$_2$ and Si. The layer number, $N$ is measured by analysing peak positions of layer breathing modes (LBMs), as measured by Raman spectroscopy.

\begin{figure*}[]
\centerline{\includegraphics[width=180mm]{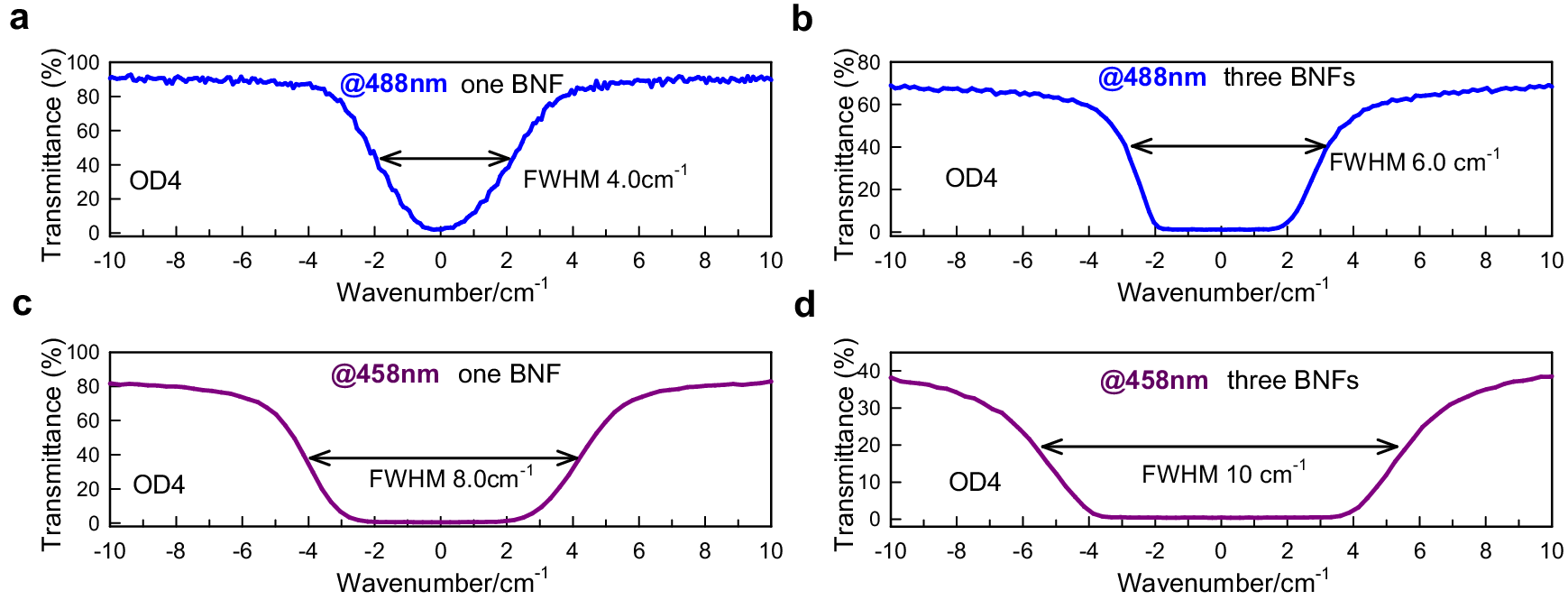}}
\caption{Transmission spectra of (a) individual and (b) three aligned VBG-based BNFs designed for 488 nm. Transmission spectra of (c) individual and (b) three aligned VBG-based designed for 458 nm.} \label{FigS1}
\end{figure*}

\textbf{Raman measurement:} Raman spectra are measured in back-scattering at room temperature with a Jobin-Yvon HR-Evolution micro-Raman system, equipped with a liquid nitrogen-cooled CCD, a $\times$100 objective lens (numerical aperture=0.90). The excitation lines at 458 nm and 488 nm are from Ar$^+$ laser. For our Jobin-Yvon HR-Evolution micro-Raman system with an 800 mm focal length, a 2048$\times$512 CCD with 13.5 $\mu$m pixel width, and a 3600 lines/mm grating, each CCD pixel covers 0.08 cm$^{-1}$ for the 488 nm laser and 0.11 cm$^{-1}$ for 458 nm. The system broadening at 488 nm is ~0.16 cm$^{-1}$, as estimated from the full width at half maximum of the Rayleigh peak. The Plasma lines are removed by using BragGrate Bandpass filters. Measurements down to $\sim$3 cm$^{-1}$ (488 nm) and 5 cm$^{-1}$ (458 nm) are enabled by three BragGrate notch filters (OptiGrate Corp) with an optical density of 4 and full width at half maximum of $\sim$6 cm$^{-1}$ and 10 cm$^{-1}$, respectively (Fig.\ref{FigS1}). The typical laser power is $\sim$ 0.3 mW to avoid sample heating. The LBM spectra are measured under ($\sigma^+$, $\sigma^+$) polarization configurations\cite{chen-nl-2015}.
%For helicity-resolved Raman measurements, a polarizer was placed in the incident path to guide the incident laser. A quarter-wave plate is inserted in the common optical path of incident and scattered light to achieve $\sigma^+$ circular polarizations. The scattered Raman signal was collected and analyzed with a half-wave plate and a linear polarizer, whose polarization is parallel to that of the polarizer in the incident optical path. Rotation of the half-wave plate at different angles enables us to obtain helicity of the scattered light, i.e., $\sigma^+$ or $\sigma^-$. The corresponding polarization configurations are denoted as ($\sigma^+$, $\sigma^+$) and ($\sigma^+$, $\sigma^-$).

\section{Interlayer bond polarizability model for LBMs in $N$L-WS$_2$}

\begin{figure*}[]
\centerline{\includegraphics[width=150mm]{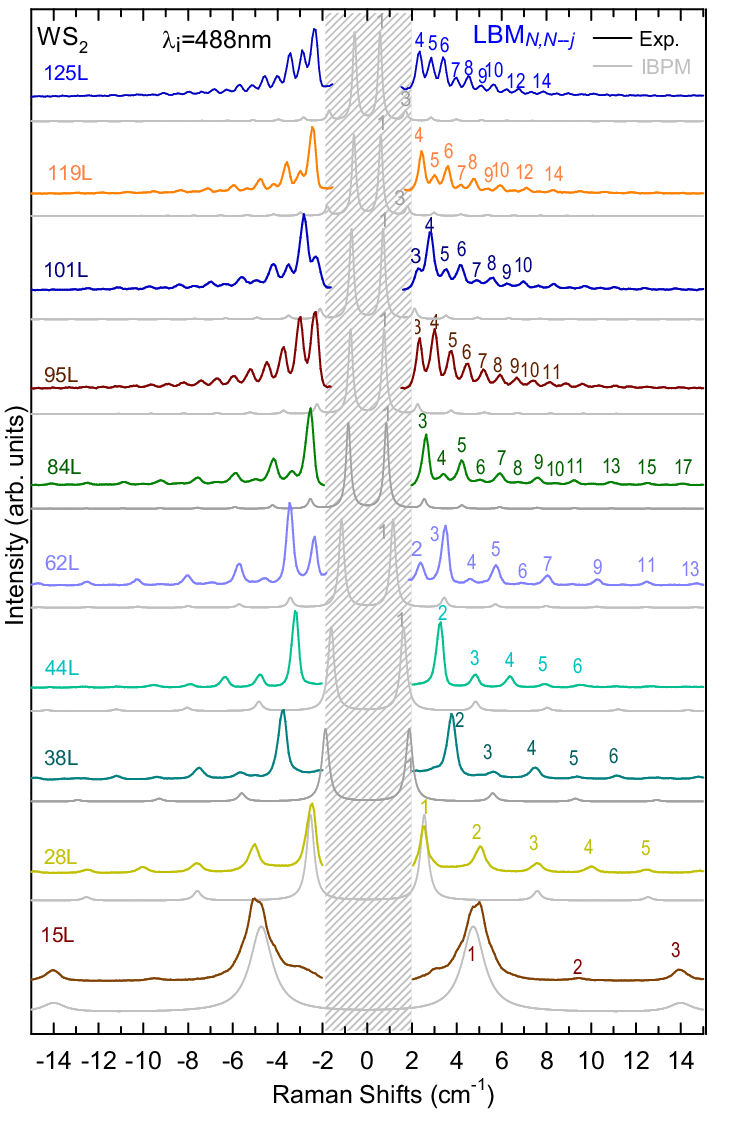}}
\caption{$N$-dependent ($N$=15-125) LBM spectra in WS$_2$ flakes (colored lines), along with the corresponding IBPM-predicted spectra (gray lines). } \label{FigS2}
\end{figure*}

In the interlayer bond polarizability model (IBPM), each layer is considered individually as the Raman intensity of LBMs, $I$(LBM), is related to the interlayer bond polarizability and bond vector\cite{luo2015stacking,liang-ns-2017}. The total change of polarizability by the interlayer vibrations of a $N$L-$MX_2$ flake is given by the sum of the changes of each layer\cite{luo2015stacking,liang-ns-2017}, $i.e., \Delta\alpha=\sum_i\alpha'_i\cdot u_i$, where $\alpha'_i$ is the polarizablity derivative of the entire layer $i$ with respect to the displacement along the $c$ axis, $u_i$ is the out-of-plane atomic plane displacement of layer $i$ calculated from the linear chain model for a specific LBM. $I$(LBM$_{N,N-j}$) is proportional to $\frac{[1+n(q_j)]}{{\rm Pos}({\rm LBM}_{N,N-j})}|\Delta\alpha|^2$, where $n(q_j)$ is the Bose$-$Einstein population factor. The IBPM\cite{luo2015stacking,liang-ns-2017} predicts that odd LBM branches can be observed in the Raman spectra, in good agreement with the analysis based on group-theory\cite{Ribeiro-prb-2014,pizzi-acsnano-2021}, as demonstrated by gray lines in Fig.1d and in Fig.\ref{FigS2}. The calculated results (gray lines in Fig. 1d and Fig.\ref{FigS2}) do not match the experimental ones when $N>10$ (blue lines in Fig. 1d and color lines in Fig.\ref{FigS2}). Thus, the IBPM alone is not enough to understand the LBM spectra of $N$L-WS$_2$ ($N>10$) under resonant excitation.

\section{Localized electronic states and electron-phonon coupling for LBMs in $N$L-WS$_2$}
\begin{figure*}[tb!]
\centerline{\includegraphics[width=180mm]{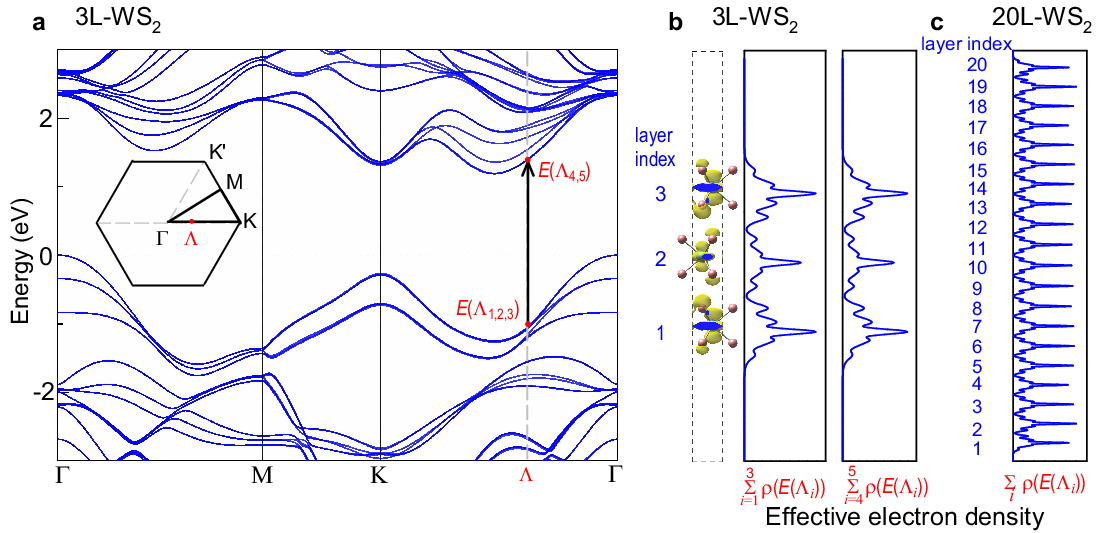}}
\caption{(a). Electronic band structure of 3L-WS$_2$. The black arrow represent the electronic transitions when the incident photon is in resonance with C exciton. The inset shows the corresponding Brillouin zone. Effective electron density related to the C exciton electronic transitions\cite{zhao2012evolution,tan2017observation} of (b) 3L-WS$_2$, and (c) 20L-WS$_2$, by summarizing the electron density ($i.e.$ square of electronic wave function) of nearly degenerate bands at valence band and conduction band.} \label{FigS3}
\end{figure*}
As the interaction between photons and standing-wave phonons within LSMs is mediated by the electronic states resonance with C excitons\cite{zhao2012evolution,tan2017observation}, we calculate the corresponding electronic states by density functional theory (DFT), as implemented in the Vienna Ab Initio Simulation Package (VASP)\cite{Kresse-PRB-1996}. The pseudopotentials are generated using the projector augmented wave (PAW) method\cite{Blochl-PRB-1994} with Perdew-Burke-Ernzerhof realization\cite{Perdew-PRL-1996} for the exchange-correlation functional. The energy cutoff is set at 350 eV for the plane-wave basis, and a $\Gamma$-centered Monkhorst-Pack $k$-point mesh with a size 21$\times$21$\times$1 is employed for the integration. A vacuum layer with a thickness$>$20 {\AA} is used, to avoid interactions between periodic images. The convergence criterion is chosen as 10$^{-5}$ eV for self-consistent electronic minimization, and 0.01 eV/{\AA} for the forces during ionic relaxation. The Van der Waals interaction is taken into account by using DFT-D3\cite{grimme-jcp-2010}. Fig.~\ref{FigS3}a plots the electronic band structure of 3L-WS$_2$.

Using lasers at 458 nm and 488 nm near the C exciton\cite{zhao2012evolution,tan2017observation}, the optical transitions occur between the density of states peaks in valence and conduction bands near the $\Gamma$ point\cite{zhao2012evolution,qiu-prl-2013}, denoted as $\Lambda$ here. As there are 3 and 2 nearly degenerate bands at valence and conduction bands, we summarize the electron density ($i.e.$ square of electronic wave functions) of these and define it as the effective electron density. The corresponding effective electron density in Fig.~\ref{FigS3}b shows that this is mainly localized at the Mo atom of each layer, around equal for each layer. A similar effective electron density distribution is seen for 20L-WS$_2$ in Fig.~\ref{FigS3}c. Thus, the electronic states related to the C excitons can be considered as one-dimensional periodic states along the $c$ axis. $M_{ e{\rm -phn}}$ in WS$_2$ should be described by the deformation-potential interaction related to delocalized LBMs and an ensemble of confined electronic states\cite{kittel1976introduction,Mlayah-prb-2007}, i.e., a spatial modulation of the crystal periodic potential by the layer displacements of the LBMs:

\begin{equation}
M_{ e{\rm -phn}}(q_jz)\propto\frac{[\frac{1}{2}\pm\frac{1}{2}+n(q_j)]}{{\rm Pos}({\rm LBM}_{N,N-j})}\int|\Psi(z)|^2 m_{\rm ph}(q_jz)dz
\label{e-phn}
\end{equation}

\noindent with $\Psi(z)$ the one-dimensional periodic electronic wave function, $m_{\rm ph}(q_jz)=\frac{\Delta u_{\rm LBM}(q_jz)}{\Delta z}$ the lattice dilation along the $c$ direction, and $n(q_j)$ the Bose-Einstein population factor. $\pm$ stand for Stokes and anti-Stokes scattering. Within the electric dipole approximation, $M_{e{\rm -pht(i/s)}}$ is considered as constant with polarization direction of incident (scattered) light. In this case, the LBM intensity is proportional to $|M_{ e{\rm -phn}}|^2$. The corresponding calculated LBM spectrum (Fig.~\ref{FigS3}) does not have even LBMs for 20L-WS$_2$, in agreement with IBPM, but not with experiments.

\begin{figure*}[tb!]
\centerline{\includegraphics[width=180mm]{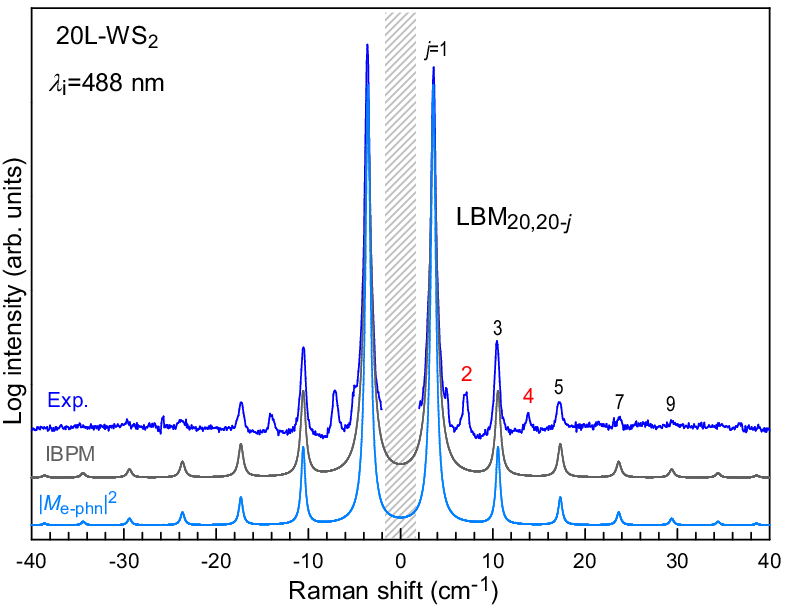}}
\caption{Experimental (blue) Raman spectra of 20L-WS$_2$, along with those predicted by the IBPM (gray) and $|M_{ e{\rm -phn}}|^2$ (light blue).} \label{FigS4}
\end{figure*}

\section{Optical cavity effect in LSMs}

We model our system as an air/$N$L-$MX_2$/substrate with complex refractive indexes $\tilde{n}_0$/$\tilde{n}_1$/$\tilde{n}_2$. The mismatches of complex refractive indices of $N$L-$MX_2$/air and $N$L-$MX_2$/substrate interfaces lead to partial reflection and transmission of incident(i) and scattered(s) photons at both top and bottom LSM interfaces, as for Fig. 2b. A superposition of two optical wave components propagating forward (down) and backward (up) (whose electric field components are denoted as $E^{\rm D}_{\rm i(s)}$ and $E^{\rm U}_{\rm i(s)}$) results in a considerable spatial modulation in the modulus square of the electric field for the incident laser and scattered light in the LSM, thus a spatial modulation for $M_{e{\rm -pht(i)}}$ and $M_{e{\rm-pht(s)}}$, as they are proportional to the vector potential of incident and scattered radiation\cite{Mlayah-prb-2007}. The combination of incident and scattered electric field components can lead to two internal scattering geometries within the LSM flake, i.e., forward (FS) and backward scattering (BS). The e-pht coupling at position $z$ [$m_{e{\rm -pht(i)}}(z) m_{e{\rm-pht(s)}}(z)$] is derived by considering the product of incident and scattered light, which can be obtained using the transfer matrix method\cite{ye-pt-1990,Li-2015-Nanoscale}:

\begin{widetext}
\begin{equation}
\begin{aligned}
m_{e{\rm-pht(i)}}(z)m_{e{\rm-pht(s)}}(z)&\propto(E^{\rm D}_{\rm i}+E^{\rm U}_{\rm i})\times(E^{\rm D}_{\rm s}+E^{\rm U}_{\rm s})\\
&=\left[\frac{t_{01}e^{{\rm i}k_{\rm i}z}}{1-r_{12}r_{10}e^{{\rm 2i}k_{\rm i}Nt}}+\frac{r_{12}t_{01}e^{{\rm i}k_{\rm i}(2Nt-z)}}{1-r_{12}r_{10}e^{{\rm 2i}k_{\rm i}Nt}}\right]\times\left[\frac{t_{10}e^{{\rm i}k_{\rm i}z}}{1-r_{12}r_{10}e^{{\rm 2i}k_{\rm i}Nt}}+\frac{r_{12}t_{10}e^{{\rm i}k_{\rm i}(2Nt-z)}}{1-r_{12}r_{10}e^{{\rm 2i}k_{\rm i}Nt}}\right]\\
&=\frac{2t_{01}t_{10}r_{12}e^{2{\rm i}k_{\rm i}Nt}}{(1-r_{12}r_{10}e^{2{\rm i}k_{\rm i}Nt})^2}+\frac{t_{01}t_{10}[e^{2{\rm i}k_{\rm i}z}+r^2_{12}e^{2{\rm i}k_{\rm i}(2Nt-z)}]}{(1-r_{12}r_{10}e^{2{\rm i}k_{\rm i}Nt})^2}\\
&=m_{e{\rm-pht}}^{\rm FS}(2k_{\rm i})+m_{e{\rm-pht}}^{\rm BS}(2k_{\rm i}z)
\end{aligned}
\label{eq:fsbs}
\end{equation}
\end{widetext}

\noindent where $k_{\rm i}=\frac{2\pi n_1}{\lambda_{\rm i}}$ is the wave vector of incident light of wavelength $\lambda_{\rm i}$ within the $N$L-$MX_2$, $t_{\mu\nu}=\frac{2\tilde{n}_{\mu}}{\tilde{n}_{\mu}+\tilde{n}_{\nu}}$ and $r_{\mu\nu}=\frac{\tilde{n}_{\mu}-\tilde{n}_{\nu}}{\tilde{n}_{\mu}+\tilde{n}_\nu}$ are the amplitude transmission and reflection coefficient from medium $\mu$ to $\nu$, respectively, $t$ is the thickness of 1L-MX$_2$. If the substrate consists of two layers, such as SiO$_2$/Si with refractive index $\tilde{n}_2/\tilde{n}_3$, $r_{12}$ should be the effective amplitude reflection coefficient $R_{12}$, i.e., $R_{12}$=$\frac{r_{12}+r_{23}e^{{\rm 2i}k_{\rm i}d}}{1+r_{12}r_{23}e^{{\rm 2i}k_{\rm i}d}}$, with $d$ is the SiO$_2$ thickness.
\section{Spatial interference model of photon-phonon coupling}
\begin{figure*}[]
\centerline{\includegraphics[width=180mm]{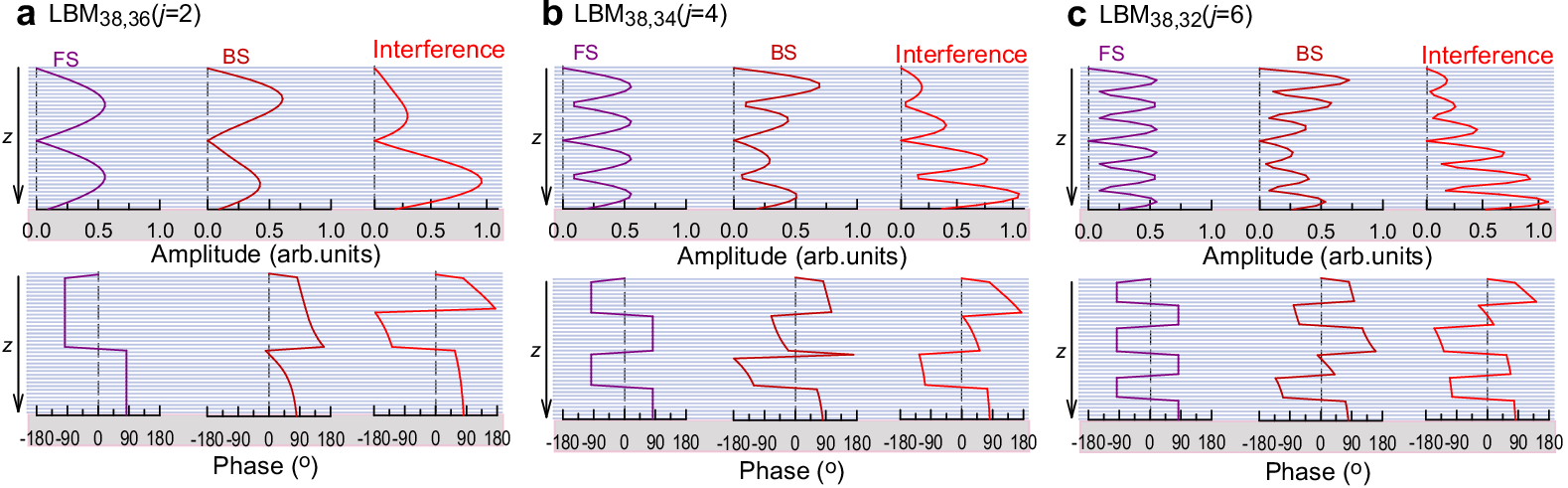}}
\caption{Spatial amplitude and phase of FS, BS and FS-BS interfering pht-phn coupling matrices for (\textbf{a}) LBM$_{38,36}$, (\textbf{b}) LBM$_{38,34}$ and (\textbf{c}) LBM$_{38,32}$} \label{FigS5}
\end{figure*}

\begin{figure*}[]
\centerline{\includegraphics[width=180mm]{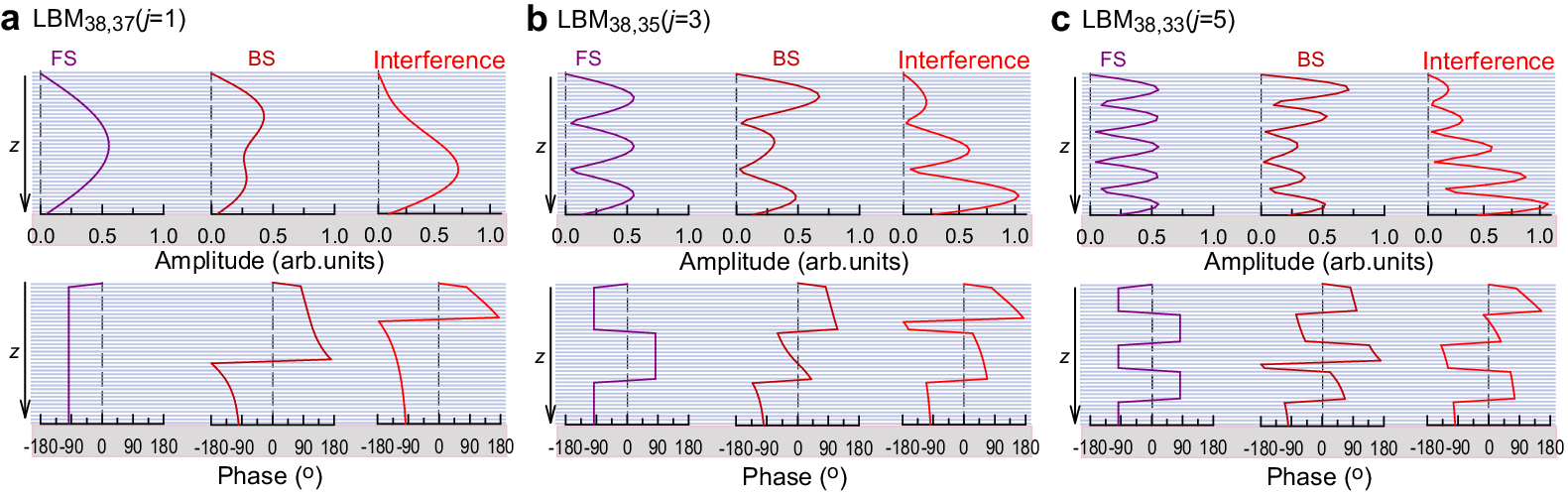}}
\caption{Spatial amplitude and phase of FS, BS and FS-BS interfering pht-phn coupling matrices for (\textbf{a}) LBM$_{38,37}$, (\textbf{b}) LBM$_{38,35}$ and (\textbf{c}) LBM$_{38,33}$.} \label{FigS6}
\end{figure*}

$M_{e{\rm-phn}}$ in LSMs is given by the deformation-potential interaction related to delocalized LBMs and an ensemble of confined electronic states, i.e., the one-dimensional periodic electronic state, as for Eq.\ref{e-phn}. The LBM resonances are defined by the phonon cavity, where the wavevectors $q_j$ of adjacent LBM resonances show a relative $\delta q$ shift. Considering the spatially coherent coupling of photons and phonons to the one-dimensional periodic electronic states in LSM, $I$($q_j$) can be obtained by a coherent summation of $[m_{e{\rm -pht}}^{\rm FS}(2k_{\rm i})+ m_{e{\rm-pht}}^{\rm BS}(2k_{\rm i}z)]m_{e{\rm -phn}}(q_jz)$ at each $z$\cite{Li-2015-Nanoscale}:

\begin{widetext}
\begin{equation}
\begin{aligned}
I(q_j)\propto &\frac{[\frac{1}{2}\pm\frac{1}{2}+n(q_j)]{\rm sin}^2(\frac{q_j t}{2})}{{\rm Pos}({\rm LBM}_{N,N-j})}\bigg|\int^{Nt}_0|\Psi(z)|^2(e^{{\rm i}q_j z}-e^{-{\rm i}q_j z})\\
&\times\left\{\frac{2t_{01}t_{10}r_{12}e^{2{\rm i}k_{\rm i}Nt}}{(1-r_{12}r_{10}e^{2{\rm i}k_{\rm i}Nt})^2}+\frac{t_{01}t_{10}[e^{2{\rm i}k_{\rm i}z}+r^2_{12}e^{2{\rm i}k_{\rm i}(2Nt-z)}]}{(1-r_{12}r_{10}e^{2{\rm i}k_{\rm i}Nt})^2}\right\}dz\bigg|^2,
\end{aligned}
\label{eq:sinc1}
\end{equation}
\end{widetext}

\noindent The integral is from the 1$^{st}$ layer to the $N^{th}$ layer, which can be considered as a summarization of the $N$ layers due to the localization of the electronic state within the layer, and then the Eq.\ref{eq:sinc1} can be rewritten as:

\begin{widetext}
\begin{equation}
\begin{aligned}
I(q_j)\propto &\frac{[\frac{1}{2}\pm\frac{1}{2}+n(q_j)]{\rm sin}^2(\frac{q_j t}{2})}{{\rm Pos}({\rm LBM}_{N,N-j})}\bigg|\sum_{n=0}^{N-1}\int^{(n+1)t}_{nt}|\Psi(z)|^2(e^{{\rm i}q_j z}-e^{-{\rm i}q_j z})\\
&\times\left\{\frac{2t_{01}t_{10}r_{12}e^{2{\rm i}k_{\rm i}Nt}}{(1-r_{12}r_{10}e^{2{\rm i}k_{\rm i}Nt})^2}+\frac{t_{01}t_{10}[e^{2{\rm i}k_{\rm i}z}+r^2_{12}e^{2{\rm i}k_{\rm i}(2Nt-z)}]}{(1-r_{12}r_{10}e^{2{\rm i}k_{\rm i}Nt})^2}\right\}dz\bigg|^2
\end{aligned}
\label{eq:sinc2}
\end{equation}
\end{widetext}

\noindent where $n$ is the layer index. As we introduced in the Supplementary Section III, the electronic states related to the C exciton can be considered as one-dimensional periodic states along the $c$ axis, $\Psi(z+nt)=\Psi(z)$. Also, considering the discrete atomic plane displacements of each layer for LBMs, the Eq.\ref{eq:sinc2} can be written as:

\begin{widetext}
\begin{equation}
\begin{aligned}
I(q_j)&\propto\frac{[\frac{1}{2}\pm\frac{1}{2}+n(q_j)]{\rm sin}^2(\frac{q_j t}{2})}{{\rm Pos}({\rm LBM}_{N,N-j})}\bigg|\int^t_0|\Psi(z)|^2dz\bigg|^2\bigg|\sum_{n=0}^{N-1}(e^{{\rm i}q_j nt}-e^{-{\rm i}q_j nt})\\
&\times\left\{\frac{2t_{01}t_{10}r_{12}e^{2{\rm i}k_{\rm i}Nt}}{(1-r_{12}r_{10}e^{2{\rm i}k_{\rm i}Nt})^2}+\frac{t_{01}t_{10}[e^{2{\rm i}k_{\rm i}nt}+r^2_{12}e^{2{\rm i}k_{\rm i}(2Nt-nt)}]}{(1-r_{12}r_{10}e^{2{\rm i}k_{\rm i}Nt})^2}\right\}\bigg|^2\\
&\propto\frac{[\frac{1}{2}\pm\frac{1}{2}+n(q_j)]{\rm sin}^2(\frac{q_j t}{2})}{{\rm Pos}({\rm LBM}_{N,N-j})}\bigg|\sum_{n=0}^{N-1}\left\{\frac{2t_{01}t_{10}r_{12}e^{2{\rm i}k_{\rm i}Nt}}{(1-r_{12}r_{10}e^{2{\rm i}k_{\rm i}Nt})^2}+\frac{t_{01}t_{10}[e^{2{\rm i}k_{\rm i}nt}+r^2_{12}e^{2{\rm i}k_{\rm i}(2Nt-nt)}]}{(1-r_{12}r_{10}e^{2{\rm i}k_{\rm i}Nt})^2}\right\}\times(e^{{\rm i}q_j nt}-e^{-{\rm i}q_j nt})\bigg|^2\\
&=\frac{[\frac{1}{2}\pm\frac{1}{2}+n(q_j)]{\rm sin}^2(\frac{q_j t}{2})}{\rm Pos({\rm LBM}_{N,N-j})}\bigg|\sum_{n=0}^{N-1}\frac{2t_{01}t_{10}r_{12}e^{2{\rm i}k_{\rm i}Nt}}{(1-r_{12}r_{10}e^{2{\rm i}k_{\rm i}Nt})^2}\times(e^{{\rm i}q_j nt}-e^{-{\rm i}q_j nt})\\
&+\frac{t_{01}t_{10}[e^{{\rm i}(2k_{\rm i}+q_j)nt}-e^{{\rm i}(2k_{\rm i}-q_j)nt}]}{(1-r_{12}r_{10}e^{2{\rm i}k_{\rm i}Nt})^2}+\frac{t_{01}t_{10}r^2_{12}e^{4{\rm i}k_{\rm i}Nt}[e^{{\rm i}(-2k_{\rm i}+q_j)nt}-e^{{\rm i}(-2k_{\rm i}-q_j)nt}]}{(1-r_{12}r_{10}e^{2{\rm i}k_{\rm i}Nt})^2}\bigg|^2\\
&=\frac{[\frac{1}{2}\pm\frac{1}{2}+n(q_j)]{\rm sin}^2(\frac{q_j t}{2})}{{\rm Pos}({\rm LBM}_{N,N-j})}\bigg|\sum_{m=0}^{N-1}\{S_{\rm FS}(2k_{\rm i},q_jn)+S_{\rm BS}[(2k_{\rm i} \pm q_j)n]\}\bigg|^2.
\end{aligned}
\label{eq:sinc3}
\end{equation}
\end{widetext}

\begin{figure*}[]
\centerline{\includegraphics[width=180mm]{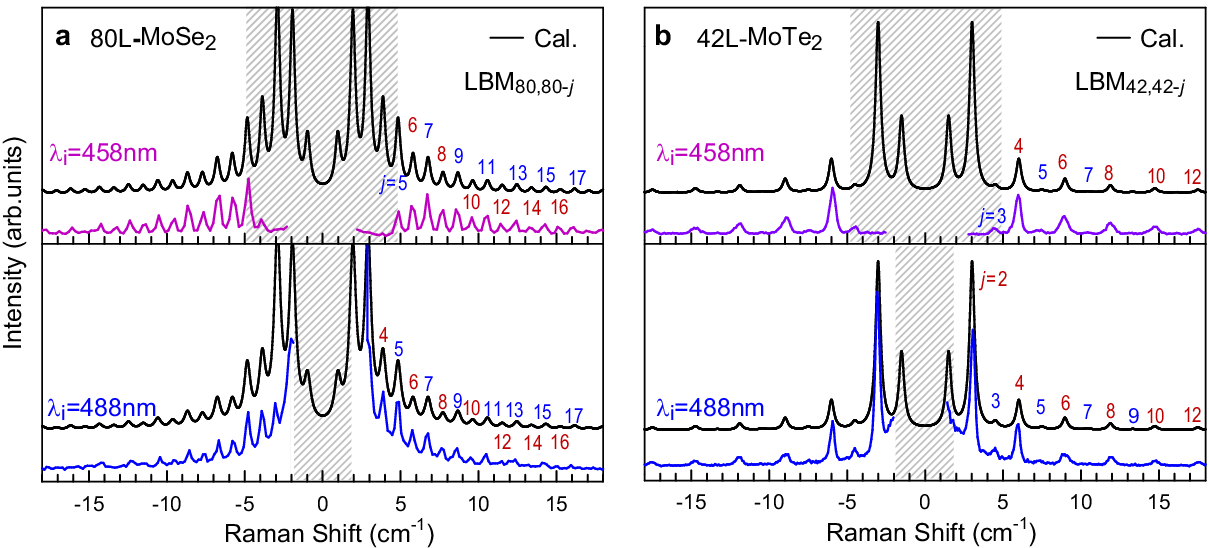}}
\caption{Experimental (colored lines) and SIM-calculated (black lines) Raman spectra for (\textbf{a}) 80L-MoSe$_2$ and (\textbf{b}) 42L-MoTe$_2$ on SiO$_2$/Si at $\lambda_{\rm i}$=458 nm and 488 nm. } \label{FigS7}
\end{figure*}

Due to the small ($<5\%$) contribution of $\frac{[\frac{1}{2}\pm\frac{1}{2}+n(q_j)]{\rm sin}^2(\frac{q_j t}{2})}{{\rm Pos}({\rm LBM}_{N,N-j})}$ to $I$($q_j$) at 0$\sim$20 cm$^{-1}$ (-20$\sim$0cm$^{-1}$) for Stokes (anti-Stokes) at room temperature, we neglect this term for simplicity. The calculated spatial variation in amplitudes and phases of $S_{\rm FS/BS}$ along $c$ axis for LBM$_{38,38-j}$ ($j$=2,4,6) and LBM$_{38,38-j}$ ($j$=1,3,5) of 38L-WS$_2$ on SiO$_2$/Si are depicted in Fig.~\ref{FigS5} and Fig.~\ref{FigS6}, respectively. To understand the physical picture behind the different intensity distribution of all the observed LBMs in LSMs, we separately discuss the contributions from internal FS and BS components. The coherent summation in $S_{\rm FS}(2k_{\rm i},q_jn)$ is irrelevant to $k_{\rm i}$, because $S_{\rm FS}(2k_{\rm i},q_jn)$ does not contain $k_i\times nt$, it can be extracted outside the coherent summarization, so that Raman intensity from internal FS component can be simplified to:

\begin{widetext}
\begin{equation}
\begin{aligned}
I_{\rm FS}&=\bigg|\sum_{m=0}^{N-1}S_{\rm FS}(2k_{\rm i},q_jn)]\bigg|^2\\
&=\bigg|\frac{2t_{01}t_{10}r_{12}e^{2{\rm i}k_{\rm i}Nt}}{(1-r_{12}r_{10}e^{2{\rm i}k_{\rm i}Nt})^2}\bigg|^2\bigg|\sum_{n=0}^{N-1}(e^{{\rm i}q_j nt}-e^{-{\rm i}q_j nt})\bigg|^2\\
&=\bigg|\frac{2t_{01}t_{10}r_{12}e^{2{\rm i}k_{\rm i}Nt}}{(1-r_{12}r_{10}e^{2{\rm i}k_{\rm i}Nt})^2}\bigg|^2\times2H^2(q_j)\{1-{\rm cos}[q_j(N-1)t]\},
\end{aligned}
\label{eq:fs}
\end{equation}
\end{widetext}

\noindent where $H(q_j)={\rm sin}(Nq_j)/{\rm sin}(q_j)$ is the interference function\cite{cazayous2002resonant} of two counterpropagating plane-wave components of each LBM, analogous to multiple-slit diffraction\cite{cazayous2002resonant,claus-ajp-1974}. $H(q_j)$ presents side maxima at odd $j$, while it reaches zero at even $j$. Thus, odd LBM branches exhibit finite intensity, while even ones show zero intensity from FS components. By contrast, the coherent summation in $S_{\rm BS}[(2k_{\rm i} \pm q_j)n]$ of each layer is closely associated with the difference between $k_{\rm i}$ and $q_j$. The corresponding $I_{\rm BS}$ is:

\begin{widetext}
\begin{equation}
\begin{aligned}
I_{\rm BS}&=\bigg|\sum_{n=0}^{N-1}S_{\rm BS}[(2k_{\rm i} \pm q_j)n]\bigg|^2\\
&=\bigg|\sum_{n=0}^{N-1}\frac{t_{01}t_{10}[e^{2{\rm i}k_{\rm i}nt}+r^2_{12}e^{2{\rm i}k_{\rm i}(2Nt-nt)}]}{(1-r_{12}r_{10}e^{2{\rm i}k_{\rm i}Nt})^2}\times(e^{{\rm i}q_j nt}-e^{-{\rm i}q_j nt})\bigg|^2\\
&=\bigg|\sum_{n=0}^{N-1}\frac{t_{01}t_{10}[e^{{\rm i}(2k_{\rm i}+q_j)nt}-e^{{\rm i}(2k_{\rm i}-q_j)nt}]}{(1-r_{12}r_{10}e^{2{\rm i}k_{\rm i}Nt})^2}+\frac{t_{01}t_{10}r^2_{12}e^{4{\rm i}k_{\rm i}Nt}[e^{{\rm i}(-2k_{\rm i}+q_j)nt}-e^{{\rm i}(-2k_{\rm i}-q_j)nt}]}{(1-r_{12}r_{10}e^{2{\rm i}k_{\rm i}Nt})^2}\bigg|^2.
\end{aligned}
\label{eq:bs}
\end{equation}
\end{widetext}

Thus, $I_{\rm BS}$ is closely associated with $H(2k_{\rm i}\pm q_j)$ and the interference between $H(2k_{\rm i}+ q_j)$ and $H(2k_{\rm i}- q_j)$, which is not only related to $q_j$ but also to $k_i$. In this case, whether odd or even LBM branches are enhanced relies on the difference between $2k_{\rm i}$ and $q_j$. As FS and BS components are coherently superposed within the $N$L-$MX_2$ cavity, the out-going Raman intensity of LBM from the LSM surface depends on the interference between FS and BS components. This leads to excitation of both odd and even LBM branches.

\section{Tunable LBM spectra}
Tunable LBM emission determined by spatially coherent pht-phn coupling is ubiquitous in LSMs, such as 80L-MoSe$_2$ and 42L-MoTe$_2$, as for Fig.\ref{FigS7}. LBMs are observed for $\lambda_{\rm i}$=458 nm and 488 nm, while the relative Raman intensity between even and odd LBMs is sensitive to $\lambda_{\rm i}$. The calculated Raman spectra show good agreement with experimental results, validating our proposed model, which accounts for the spatially coherent pht-phn coupling for LBMs mediated by ensembles of localized electronic states in LSM cavities.

\bibliographystyle{naturemag}